\documentclass[fleqn,usenatbib]{mnras}

\usepackage{newtxtext,newtxmath}
\usepackage[T1]{fontenc}
\usepackage{ae,aecompl}

\usepackage{graphicx}	% Including figure files
\usepackage{amsmath}	% Advanced maths commands
\usepackage{amssymb}	% Extra maths symbols
\usepackage{comment}
\usepackage{subfig}
\title[GW events in {\it Gaia}]{Electromagnetic counterparts to gravitational wave events from {\it Gaia}}

\author[Z. Kostrzewa-Rutkowska et al.]{
Z. Kostrzewa-Rutkowska,$^{1,2,3}$\thanks{E-mail: zkostrzewa@strw.leidenuniv.nl}
P.G. Jonker,$^{2,3}$ S.T. Hodgkin,$^{4}$ D. Eappachen,$^{2,3}$ \newauthor
D.L. Harrison,$^{4,5}$ S.E. Koposov,$^{6}$ G. Rixon,$^{4}$ {\L}. Wyrzykowski,$^7$ A. Yoldas,$^{4}$ \newauthor E. Breedt,$^{4}$ A. Delgado,$^{4,8}$ M. van Leeuwen,$^{4}$ T. Wevers,$^{4}$ P.W. Burgess,$^{4}$ \newauthor
F. De Angeli,$^{4}$ D.W. Evans,$^{4}$ P.J. Osborne,$^{4}$ M. Riello$^{4}$ 
\\
$^{1}$Leiden Observatory, Leiden University, PO Box 9513, NL-2300 RA Leiden, the Netherlands\\
$^{2}$SRON, Netherlands Institute for Space Research, Sorbonnelaan 2, 3584 CA Utrecht, the Netherlands\\
$^{3}$Department of Astrophysics/IMAPP, Radboud University, P.O. Box 9010, 6500 GL Nijmegen, the Netherlands\\
$^{4}$Institute of Astronomy, Madingley Road, Cambridge CB3 0HA, United Kingdom\\
$^{5}$Kavli Institute for Cosmology, University of Cambridge, Madingley Road, Cambridge CB3 0HA, United Kingdom\\
$^{6}$McWilliams Center for Cosmology, Carnegie Mellon University, 5000 Forbes Ave, 15213, USA\\
$^{7}$Warsaw University Astronomical Observatory, Al. Ujazdowskie 4, 00-478 Warszawa, Poland\\
$^{8}$Centre for Astrobiology (CAB - CSIC/INTA), ESAC, Madrid, Spain 
}

\date{Accepted XXX. Received YYY; in original form ZZZ}

\pubyear{2019}

\begin{document}
\label{firstpage}
\pagerange{\pageref{firstpage}--\pageref{lastpage}}
\maketitle

\begin{abstract}
The recent discoveries of gravitational wave events and in one case also its electromagnetic (EM) counterpart allow us to study the Universe in a novel way. The increased sensitivity of the LIGO and Virgo detectors has opened the possibility for regular detections of EM transient events from mergers of stellar remnants. Gravitational wave sources are expected to have sky localisation up to a few hundred square degrees, thus {\it Gaia} as an all-sky multi-epoch photometric survey has the potential to be a good tool to search for the EM counterparts. In this paper we study the possibility of detecting EM counterparts to gravitational wave sources using the {\it Gaia} Science Alerts system. We develop an extension to current used algorithms to find transients and test its capabilities in discovering candidate transients on a sample of events from the observation periods O1 and O2 of LIGO and Virgo. For the gravitational wave events from the current run O3 we expect that about 16 (25) per cent should fall in sky regions observed by {\it Gaia} 7 (10) days after gravitational wave. The new algorithm will provide about 10 candidates per day from the whole sky.
%193
\end{abstract}


%%%%%%%%%%%%%%%%%%%%%%%%%%%%%%%%%%%%%%%%%%%%%%%%%%

%%%%%%%%%%%%%%%%% BODY OF PAPER %%%%%%%%%%%%%%%%%%

\section{Introduction}

The recent detection of an electromagnetic (EM) counterpart to the gravitational wave (GW) event GW170817 represents a major advance in multi-messenger astronomy (\citealt{2017ApJ...848L..12A}). This event was caused by the merger of two neutron stars, which resulted in a kilonova event (AT2017gfo) at 40 Mpc, visible in multi-wavelength observations (from gamma rays to radio). Several studies have been published addressing questions regarding the origin of this event, its observable properties, the physics of kilonova models, the abundances of r-process elements in the Universe, and the  cosmological implications, just to mention a few (e.g. \citealt{2017Natur.551...85A,2017Natur.551...67P,2017Natur.551...75S}). During the first two periods where the LIGO (and Virgo) GW detectors have been operational called observing runs (O1 -- from 2015 September 12 to 2016 January 19 and O2 -- from 2016 November 30 to 2017 August 25), eleven GW events have been observed by the LIGO and Virgo detectors (\citealt{2019PhRvX...9c1040A}). Ten of these were associated with black hole - black hole (BHBH) mergers and one with a neutron star - neutron star (NSNS) merger and its kilonova signal as mentioned above (\citealt{2016PhRvL.116x1103A,2017PhRvL.118v1101A,2017ApJ...848L..12A,2017PhRvL.119n1101A}). Black hole - neutron star (BHNS) mergers have remained undetected during these runs. 

Predictions are that up to 50 events could be detected from NSNS mergers and BHNS mergers after the improved detectors have been operational for one year (\citealt{2018PhRvL.120i1101A}). This third observing run (O3) started on 2019 April 01. Several papers have been published predicting that one might be also able to observe an EM counterpart to BHBH mergers (e.g. \citealt{2017ApJ...839L...7D,2017PhRvD..96l3003K}). The brightness of kilonova events is estimated at 19.5 and 21 mag in optical bands at 100 and 200 Mpc, respectively (\citealt{2017LRR....20....3M}). Candidate events for BHBH, NSNS, and BHNS mergers have been already reported in this run (e.g. \citealt{2019GCN.24098....1L,2019GCN.24168....1L,2019GCN.25333....1L}).

The ESA-{\it Gaia} mission has been operational since mid-2014 and has provided accurate photometric, astrometric, and spectroscopic measurements for roughly a billion stars in the Milky Way (\citealt{2018A&A...616A...1G}). {\it Gaia}'s on-board detection algorithms are optimised for the detection of point-like sources, although the mission is also collecting data for a significant number of resolved extragalactic objects (\citealt{2014sf2a.conf..421D}). As a result of the observing strategy, {\it Gaia} scanned over the location of most of the sources more than 70 times from different angles during the first five year mission. Each position on the sky is observed, on average, once every 30 days (\citealt{2016A&A...595A...4L}). These repeat visits make {\it Gaia} an all-sky, multi-epoch photometric survey that allows us to monitor variability with high precision, as well as detect new transient sources (\citealt{2013RSPTA.37120239H,2017arXiv170203295E}). The Data Processing and Analysis Consortium (DPAC) handles {\it Gaia}'s data flow, and this enables the detection of transients within 24-48 hours of observations. After September 2014 new transients from {\it Gaia} have been made publicly available after manual vetting of candidate transients detected by the {\it Gaia} Science Alerts (GSA) team (see: \url{http://gsaweb.ast.cam.ac.uk/alerts/alertsindex} and \citealt{2019ASPC..521..507D,2019ASPC..523..261D}). To this end, AlertPipe - dedicated software for data processing, transient searching, and candidate filtering was employed (Hodgkin et al. in prep.). {\it Gaia}'s accurate photometry and low-resolution spectroscopy should allow for a robust classification and reduces the rate of false positives. {\it Gaia} could therefore play an important role as a transient detection survey. Several publications have shown that {\it Gaia} is able to detect uncommon transients such as the eclipsing AM~CVn system Gaia14aae (\citealt{2015MNRAS.452.1060C}), the superluminous supernova Gaia16apd (\citealt{2017MNRAS.469.1246K,2017ApJ...835L...8N}), fast transients (\citealt{2018MNRAS.473.3854W}), transients in the centres of galaxies (\citealt{2018MNRAS.481..307K}), just to mention a few. Moreover, in July 2019 {\it Gaia}'s mission entered in the extension period from mid-2019 to the end of 2020 and it is likely that the mission will be extended further with a firm end-of-mission date of end of 2024 ($\pm$6 months).

In this study we aim to explore the possibility of an improvement for the detection rate so that the GSA can capture transient events that the existing pipeline might miss (as they are too faint and/or too fast). Specifically, we propose to employ a bespoke detection algorithm for the {\it Gaia} data to search for EM counterparts to GW events. This new detection algorithm will make use of GW event localisation and timing to allow it to run at a lower detection threshold and thanks to that we will increase the completeness (although the sample purity might decrease).
We perform a systematic search for transients that coincide in time and in sky localisation with the run O1 and O2 GW detections (\citealt{2019PhRvX...9c1040A}), to investigate if a dedicated source finding and vetting algorithm can be implemented to run during (the remainder of) LIGO/Virgo's O3 and O4 so that {\it Gaia} can enhance its contribution to the search for the EM counterpart to a GW event. We also provide a list of potential EM transients that occurred close in time and sky location to GW events from the O1 and O2 runs.  

This paper is organized as follows. In Section \ref{sec:detector} we present properties of, and tests on, the new detection algorithm, and discuss our results from a one year test. In Section \ref{sec:O1O2} we show the results from a search for candidate transients coincident with GW events from the O1 and O2 runs, and furthermore, consider implications for the future {\it Gaia} possibilities of detecting the EM signal associated with GWs. We conclude in Section \ref{sec:con}.
Throughout this paper we assume a flat $\Lambda$-Cold Dark-Matter ($\Lambda$CDM) concordance cosmological model of the Universe with parameters $\Omega_\Lambda = 0.7$, $\Omega_\mathrm{M} = 0.3$ and $H_0 = 70\mathrm{~km~s^{-1}~Mpc^{-1}}$, $h = 0.70$.

\section{GW detector}
\label{sec:detector}

As with any optical transient detected by the GSA pipeline, an EM counterpart to a GW event might be found as a so called "new source", which is a source that has not previously been seen by {\it Gaia}, or as an existing source changing its photometric properties (typically brightening although the GSA pipeline also detects sources fading). A "new source" classification for instance occurs if any host galaxy is below the detection threshold or if the transient is resolved from the host galaxy light and it passes the thresholds set in the detection algorithm for a new independent source detection. The existing GSA {\it NewSource} detection algorithm, a detector in short, will trigger a detection if the event has 2 or more detections above a flux threshold equivalent to {\it Gaia}'s $G$--band magnitude of $G=19$ mag, and it is detected by observations made with both of {\it Gaia}'s telescopes (i.e.~ the source is detected in the two different fields-of-view; Hodgkin et al. in prep.).

Here, we propose an additional detector to increase the efficiency in finding new events at the expense of a higher intrinsic rate of false positives that have to be filtered out. We investigate the influence of a lower flux threshold and the removal of the condition that at least 2 observations should detect the new source has on the transient detection rate. The removal of the requirement of two {\it Gaia} detections is driven by kilonova models for the evolution of the source (absolute) magnitude with time. From those models we expect at most an optical signal detectable by {\it Gaia} for up to a week after the merger for likely distances (\citealt{2017LRR....20....3M}). Hence, given that a second scan by {\it Gaia} over the same sky location that led to the first detection might only take place after the source has faded making it potentially too faint to be detectable by {\it Gaia} implies that the current GSA system would not flag the source as a transient (specifically 25 per cent of time sources will only have one observation and a fraction of the data may be held on board until later). However, as said, removal of the requirement of the second detection will also increase the number of false positive detections. To mitigate these effects we developed and tested additional filters to verify the photometric and astrometric data accuracy. 

\begin{figure}
\includegraphics[width=\columnwidth]{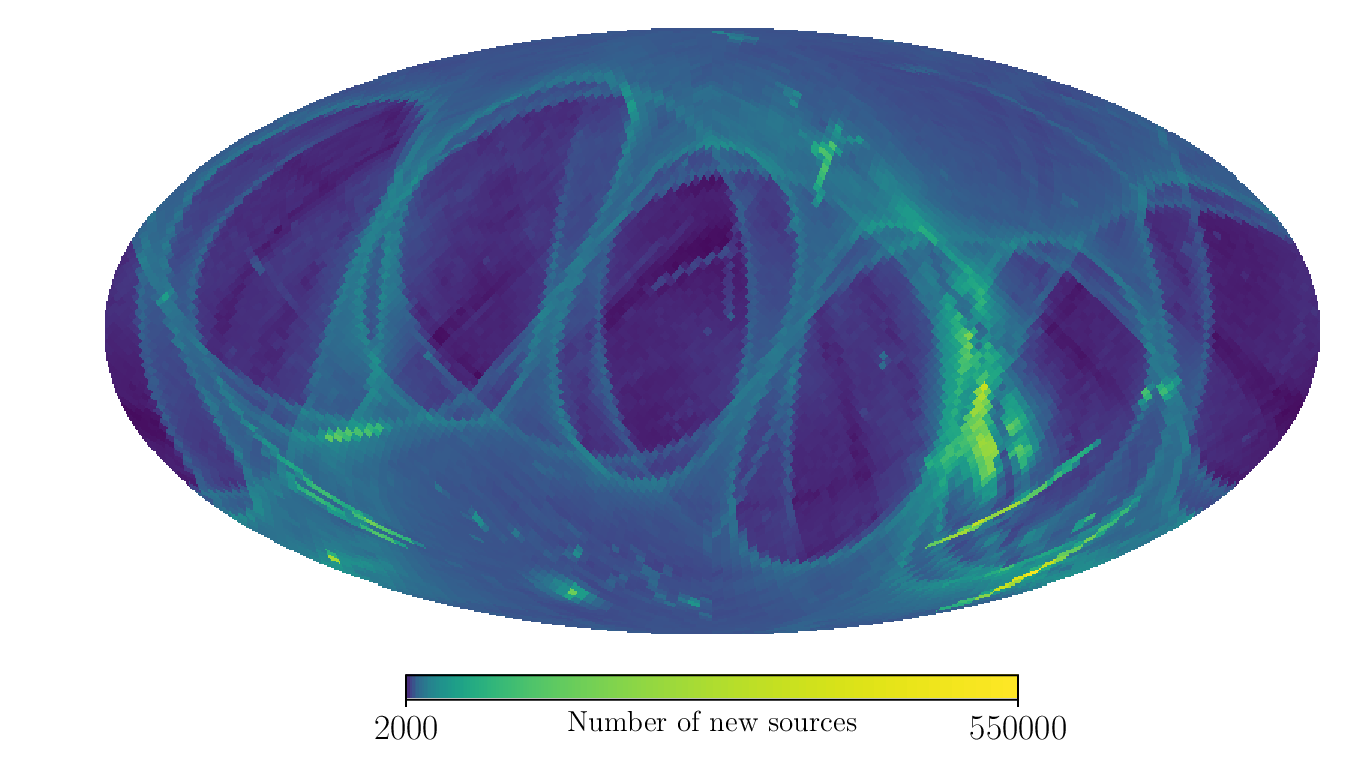}
\caption{The all-sky distribution of new sources flagged by IDT during the year 2018. The artefacts from the scanning law are clearly visible. The map resolution is HEALPix of nside 32. The plot is in equatorial coordinates, 0,0 in the centre, with North up and East to the left.}
\label{fig:densityall}
\end{figure}

In this study we make use of all {\it Gaia} photometric data collected since the beginning of the mission (July 2014) that are ingested into the {\it Gaia} Science Alerts Database (GSA DB). By using the GSA DB we have access to the {\it Gaia} time series and individual measurements from scans (=transits) as opposed to data available in {\it Gaia}'s early data releases where only averaged data products are available. To build and study properties of a new detector we have taken one year of data (from the full year of 2018) where we searched for all transits classified as new sources during the initial data treatment (IDT, \citealt{2016A&A...595A...3F}) by DPAC. A new source is created when no match is found within a radius of 1.5 arcsec. There are more than 144 million transits flagged by IDT as new sources. Fig.~\ref{fig:densityall} shows the density distribution of those sources over the sky. The map shows artefacts from {\it Gaia}'s scanning law and it shows that many of those tentative new sources fall in the most dense regions on the sky (in the Milky Way bulge and disc, and near to the Magellanic Clouds).

Due to our relaxed requirement of including sources that were detected by a single detection we will be much more susceptible to the detection of artefacts introduced by nearby bright stars, close binary systems, dense regions, planets, solar system objects (SSOs), and epochs of initially bad astrometry, to mention just a few of the effects that can cause spurious detections in {\it Gaia}'s IDT. Furthermore, our lower flux threshold for transients when compared to the existing GSA detection algorithms will give us additional samples of transients, some of which are spurious, that are usually not detected by the standard GSA system. Hence, additional cuts and filtering must be applied in order to weed out the spurious transients as much as possible. 

\subsection{Selection process}
\label{sec:selection}

\begin{figure}
\includegraphics[width=\columnwidth]{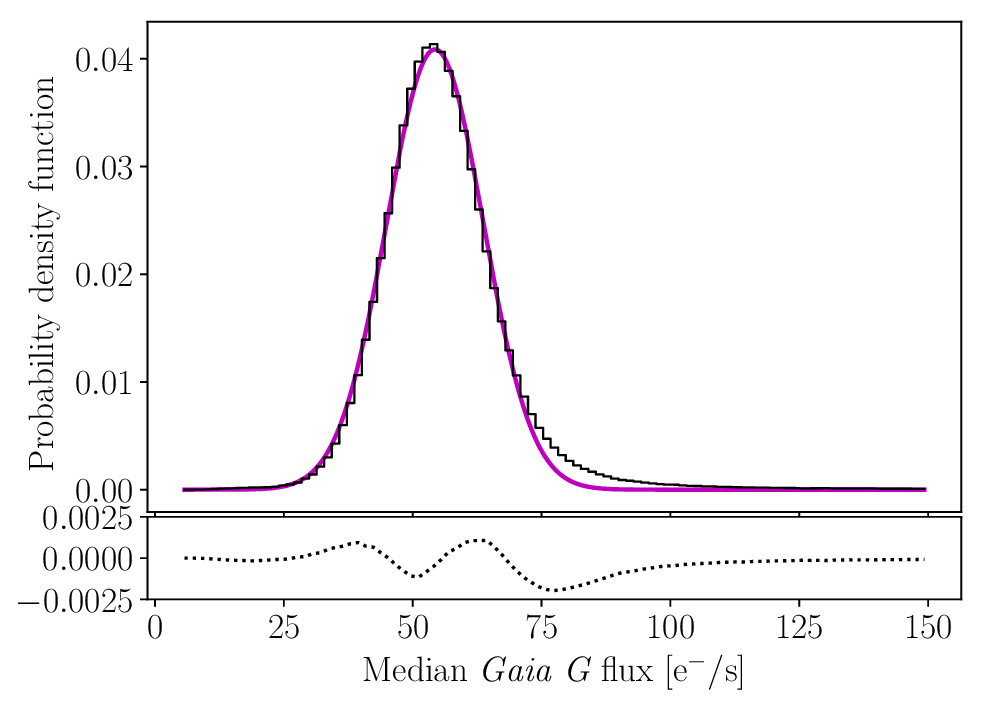}
\caption{The distribution of the median flux from single transits for new sources that were also detected in the PS1 DR1. The magenta line indicates a Gaussian fit to the distribution. We remove these sources from the list of transients detected by {\it Gaia} as they are likely caused by Poisson fluctuations (or real low-amplitude variability) in fainter sources detected previously by PS1.}
\label{fig:fluxPS1xG}
\end{figure}

\begin{figure}
\includegraphics[width=\columnwidth]{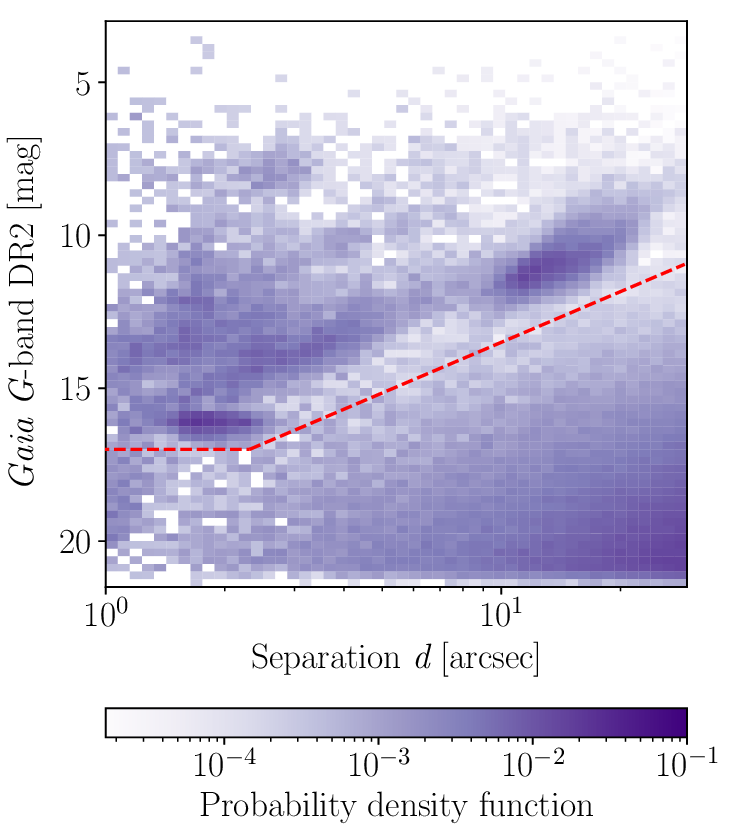}
\caption{The distance between a potential ''new source'' and its nearest neighbours found in GDR2 within a search radius of 30 arcsec around the position of the ''new source'' vs.~the $G$-band brightness of those neighbours in GDR2. A group of new candidate transients appears in close proximity of bright stars. The plot shows a sample of 1 per cent from all candidate transients detected. The red dashed line indicates the applied cut, where all candidate transients above the red dashed line are removed as they are likely to be caused by artefacts such as diffraction spikes caused by bright stars. }
\label{fig:dr2neighbour}
\end{figure}

The final selection of candidate transients in the novel detector has been performed using the following filtering steps:

\begin {enumerate}
    \item We required that at least 8 of the 9(8) astrometric field (AF) CCD measurements during a single transit must return a valid photometric data point (see \citealt{2015A&A...576A..74D} for a description of {\it Gaia}'s focal plane).

    \item We excluded the most dense regions in the sky (the bulge and disc of the Milky Way, and the Large and Small Magellanic Clouds) by applying a cut on the number of sources per HEALPixel of nside 4096 equivalent to $50\times50$ arcsec (see \citealt{2005ApJ...622..759G} for the definition of HEALPixels and their sizes). We made use of {\it Gaia} Data Release 2 (GDR2, \citealt{2018A&A...616A...1G}) source density maps and decided that the HEALPixel is excluded from the further processing if the number of sources is larger than the mean number of sources per HEALPixel. In Fig.~\ref{fig:densitycuts} we show density maps from GDR2 before and after applying the cut.
    
    \begin{figure}
\includegraphics[width=\columnwidth]{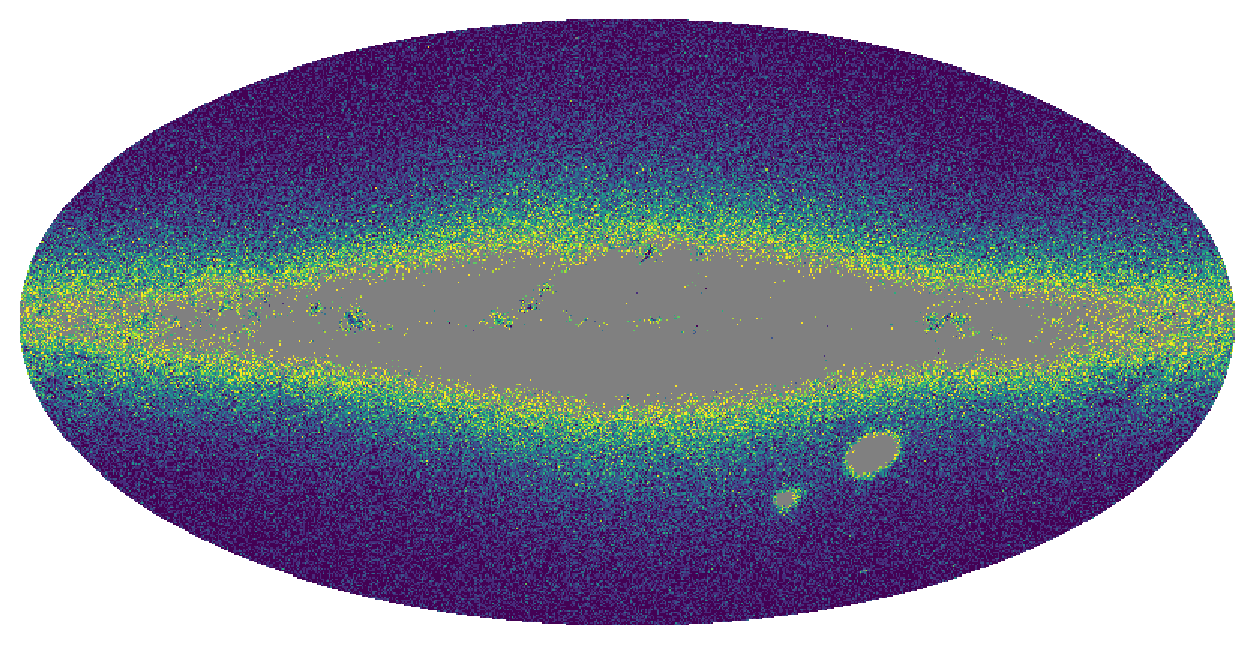}
\caption{The density map of sources in GDR2 (the number of sources per HEALPixel of nside 4096) in HEALPixels chosen for the new detector. Grey regions indicate deselected parts as they have a number of sources above the mean. The cut mostly affects fields in the Milky Way bulge and disc, and the centres of the Small and Large Magellanic Clouds. About 21 per cent of the sky was removed and will not be processed during the search for EM counterparts to GW events. The plot is in Galactic coordinates.}
\label{fig:densitycuts}
\end{figure}

    \item We removed all new sources created during astrometric excursions of the satellite (caused by e.g.~hits by micro-meteorites, space debris, and non-rigidity events, see: \citealt{2008IAUS..248...82V}). This effect may cause a significant excess of number of detected new sources during the IDT. The reason is that sources are preliminarily assigned erroneous coordinates due to the astrometric excursions, which influences the low-latency GSA DB. Further processing later in time, well before the formal {\it Gaia} Data Releases, corrects for this. Here, we created a histogram of number of observed new sources as a function of time (with a time bin size of about 20 minutes) to eliminate data collected during astrometric excursions as those stand out as peaks where the "transient" discovery rate shoots up. We noticed 6 major events where the number of new sources rose up to $10^4$ or more per 20 minute intervals and several lower peaks are present as well (see Fig. \ref{fig:obstime}). We removed all bins where the number of observed sources is larger $1 \sigma$ above the 20-minute average (i.e. about 3000 sources per 20 minutes).
   
   \begin{figure}
\includegraphics[width=\columnwidth]{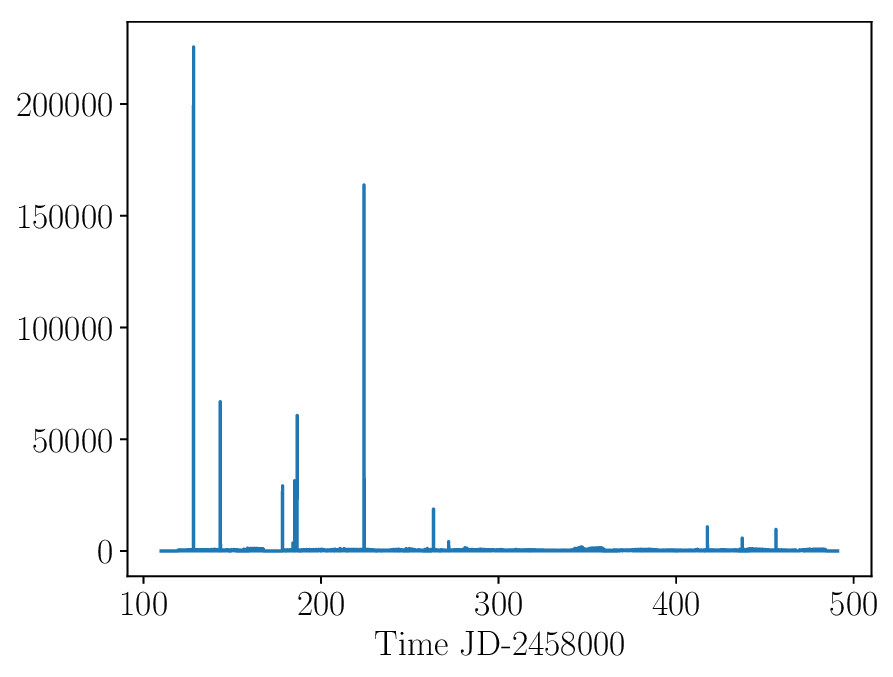}
\caption{The number of new sources created during the IDT vs.~time for the year 2018. The remarkable peaks where the number of new sources reaches up to $10^4$ and more appear a few times in the first half of the year. Scanning the Milky Way and ecliptic also causes an excess in the rate of detected transients due to problems with crowding and solar system object detections. Several smaller peaks are also visible all over the year. The epochs of these enhanced rates of detections of transients are removed (see \ref{sec:selection} for the exact threshold used).}
\label{fig:obstime}
\end{figure}
  
   \item Magnitude limit: we required the median 9(8) CCD flux to be $>101.25$ $\mathrm{e^{-}/s}$ (equivalent to $G \sim 20.68$ mag, calibrated as in the GSA pipeline, Hodgkin et al. in prep.). We got to this limit as follows: all detected transients that remained after the filters listed above were cross-matched with the Pan-STARRS Data Release 1 catalogue (PS1 DR1, \citealt{2016arXiv161205560C}) within a search radius of 1 arcsec. Many new {\it Gaia} detections coincide with known (fainter) sources from the PS1 survey. We studied the flux distribution of these sources in {\it Gaia} and fitted a Gaussian function to their flux distribution in {\it Gaia} (see Fig.~\ref{fig:fluxPS1xG}). All sources fainter than $5\sigma$ above the mean value of flux distribution, i.e.~$G \sim 20.68$ mag, are likely to be sources that are detected by {\it Gaia} and labelled during the IDT as a new source due to a Poisson fluctuation in their count rate. Therefore, setting the detection threshold at $5\sigma$ above the mean value of flux distribution of these candidate transients only a small number of these spurious sources remains (in Gaussian statistics only one in a million).
   
   \item We removed artefacts from bright stars using data from GDR2.
   Bright stars might cause multiple spurious detections in a large radius around them (\citealt{2015A&A...576A..74D}). For each candidate new source we search for all neighbours within a search radius of 30 arcsec in GDR2.
   The plot in Fig.~\ref{fig:dr2neighbour} shows the offset $d$ between the candidate and neighbours from GDR2 versus the $G$-band mean magnitude of the neighbours.
   We assumed that any source fulfilling the condition $-5.5\cdot \log(d\mathrm{[arcsec]})+19 < G$ (the diagonal red dashed line in Fig. \ref{fig:dr2neighbour}) may cause an artefact in detection. However, as the counterparts to the GW events will be located in galaxies we have to prevent the situation when the closest and the brightest neighbour is actually the centre of the host galaxy. Hence, we also studied the distribution of galaxy brightness in GDR2. Using a sample of spectroscopically confirmed galaxies from the SDSS catalogue (\citealt{2017AJ....154...28B}) we found that only 1 per cent of those detected by {\it Gaia} is brighter than $G=17$ mag (Fig.~\ref{fig:sdssgals}). Therefore, we decided not to exclude any candidate with a neighbour in GDR2 fainter than $17$ mag in $G$-band (the horizontal part of the red dashed line in Fig. \ref{fig:dr2neighbour}). 
   
   \item Scatter during transit: the median absolute deviation (MAD) within 9(8) CCD flux measurements during a single transit must be limited. We assumed no significant change in the light curves within the crossing time through 9(8) CCDs that the source transits over the focal plane (a single transit lasts about 45 seconds). In general, the scatter is a function of flux, hence the cut we apply is a function of the source brightness (see Fig.~\ref{fig:fluxerrors}).
   
   \item We removed artefacts from bad cross match during IDT (caused for example by bad astrometry) by performing an internal cross match within the GSA DB. If a candidate has neighbouring transits within $0.5$ arcsec detected before the detection of the candidate we assume that these two entries in the database should be considered as the same source (and the new source under consideration was in fact erroneously not matched during IDT to that nearby source detected before). 
   
   \item We removed transits where the photometry is flagged as bad during the IDT.

   \item We excluded all new sources potentially caused by SSOs. All new sources were cross matched with the available internal SSO table (Hodgkin et al. in prep.) within a search radius of 2 arcmin and within a time difference between observations of 3 seconds. The source and a predicted SSO transit have to be observed by the same field-of-view, on the same CCD row, within 3 seconds of time, and with a distance offset less than 0.1 arcmin (see Fig.~\ref{fig:ssoxmatch} for the distribution of offsets between new sources and SSOs).  
   
\end{enumerate}

\begin{figure}
\includegraphics[width=\columnwidth]{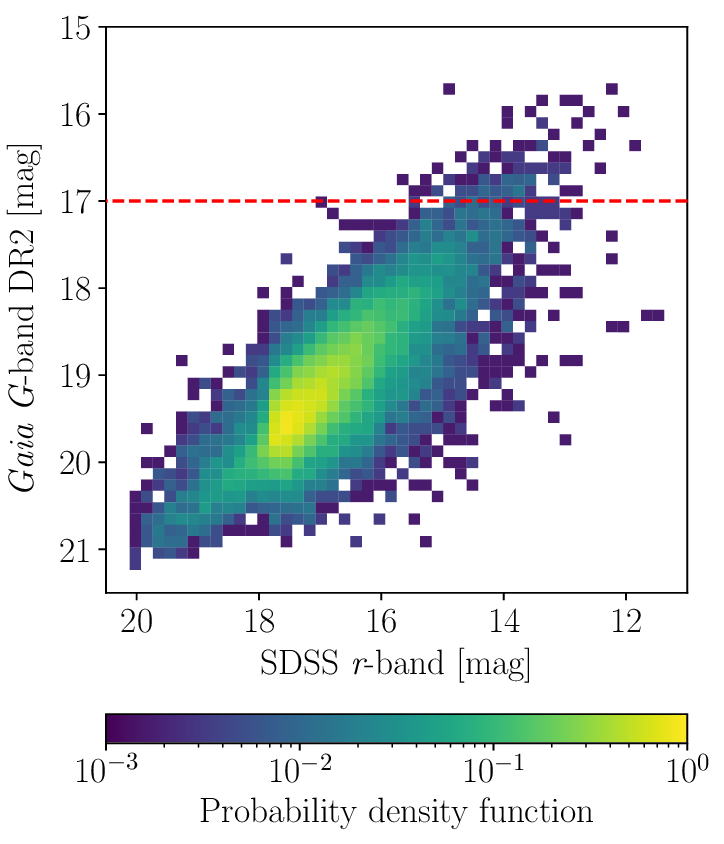}
\caption{The magnitude of the source detected by {\it Gaia} ($G$-band mean) vs. the magnitude of the SDSS source ($r$-band model magnitude) for a sample of spectroscopically confirmed galaxies from the SDSS catalogue cross matched with GDR2 using a search radius of 1 arcsec. About 25 per cent of galaxies was detected by {\it Gaia} and included in GDR2. The {\it Gaia} detections of the extended SDSS sources return typically fainter magnitudes in comparison to the SDSS brightness as {\it Gaia} only probes the central parts of the galaxies. About 1 per cent of the detected objects is brighter than $17$ mag in $G$-band (sources above the red dashed line). Hence, by  excluding any candidate new source located in the vicinity of GDR2 source brighter than $17$ mag we might remove from the sample about 1 per cent of possible transients in extended hosts.}
\label{fig:sdssgals}
\end{figure}

\begin{figure}
\includegraphics[width=\columnwidth]{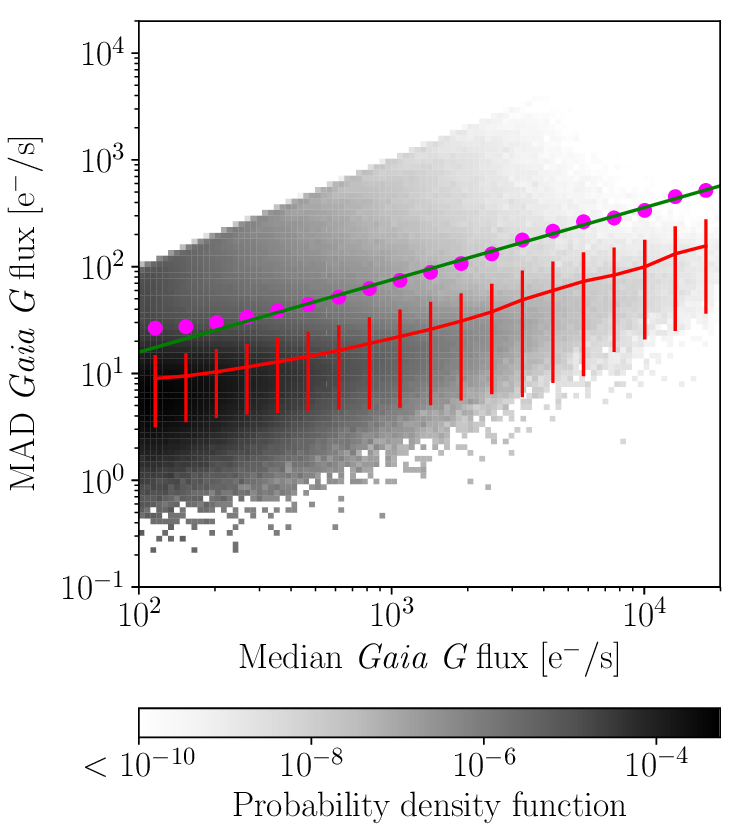}
\caption{The distribution of the median absolute deviation (MAD) vs.~the median source flux during a single transit. The red line indicates the median value of MAD in flux bins with $1\sigma$. The magenta points indicate $3\sigma$ above the median. To link the cut on scatter to the value of the flux we fitted a line (in a log-log space) to $3\sigma$ points. The best fit is shown as a green line ($\sim \mathrm{flux}^{0.68\pm0.02}$). Sources that fall above this line are excluded. }
\label{fig:fluxerrors}
\end{figure}

\begin{figure}
\includegraphics[width=\columnwidth]{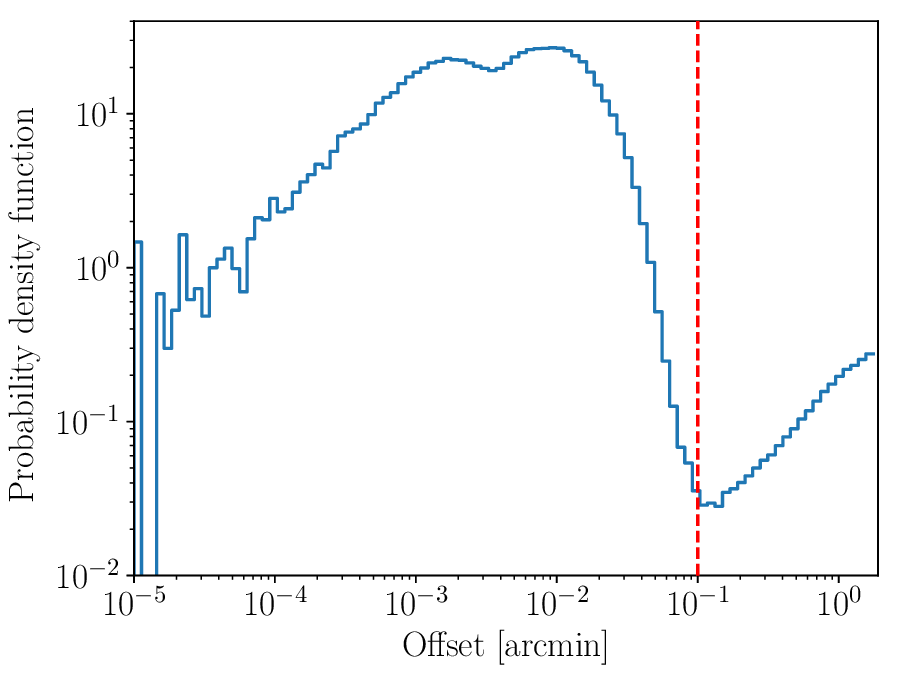}
\caption{The distribution of coordinate offsets between candidate transients and SSOs found within 2 arcmin. The time span between a candidate observation and predicted time of SSO observation is lower than 3 seconds. We removed candidate transients found within 0.1 arcmin (all object to the left of the red dashed line) and 3 seconds of an SSO.}
\label{fig:ssoxmatch}
\end{figure}

\begin{figure}

\subfloat[]{\includegraphics[width=0.9\columnwidth]{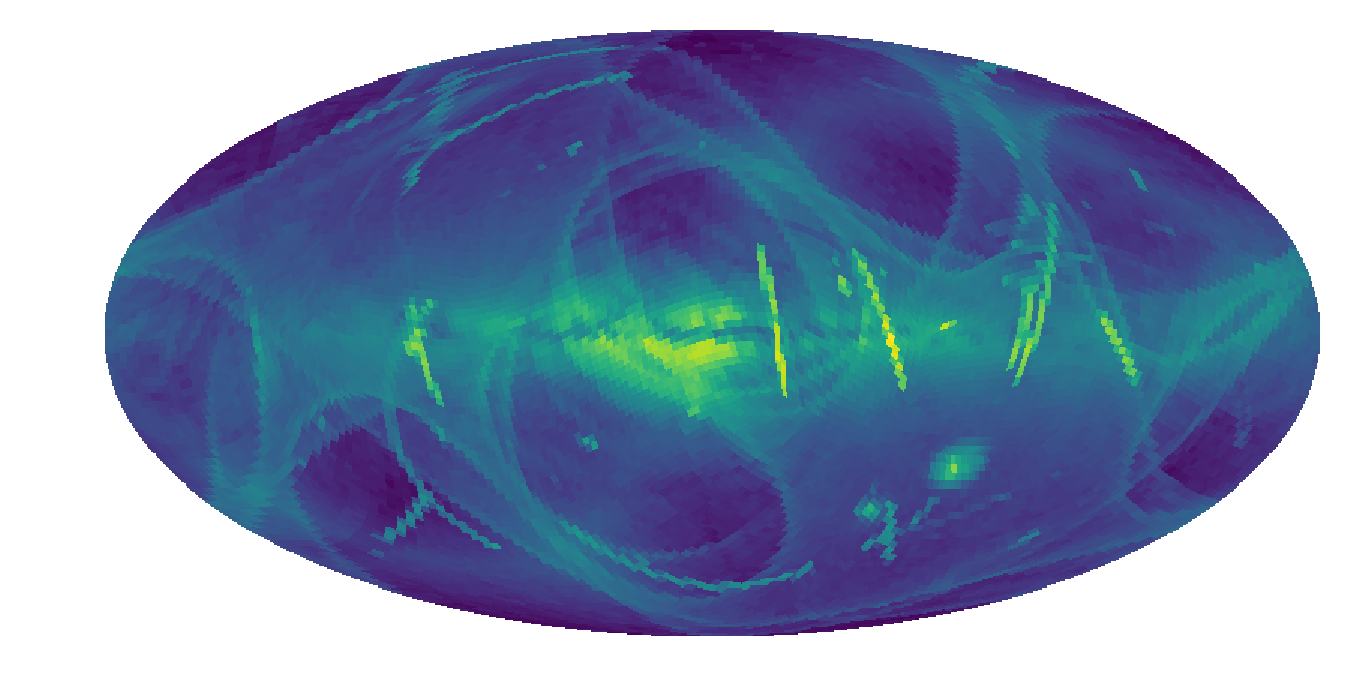}} 

\subfloat[]{\includegraphics[width=0.9\columnwidth]{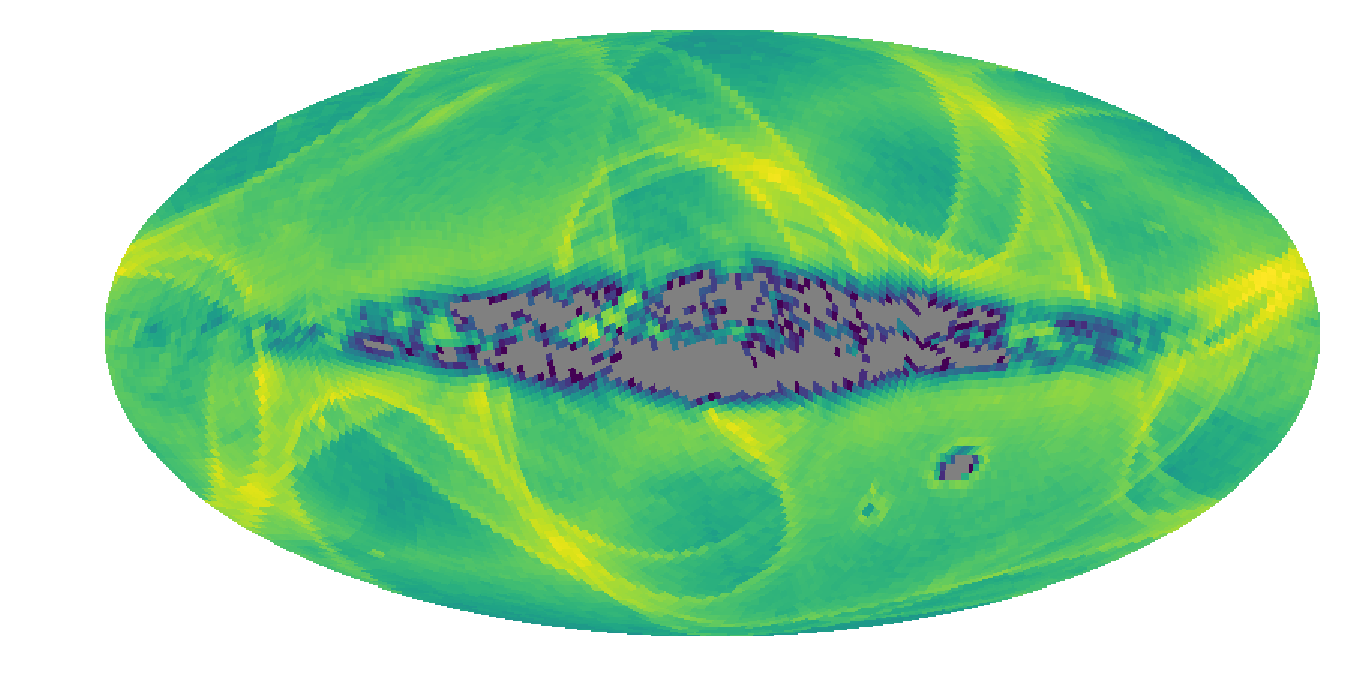}}

\subfloat[]{\includegraphics[width=0.9\columnwidth]{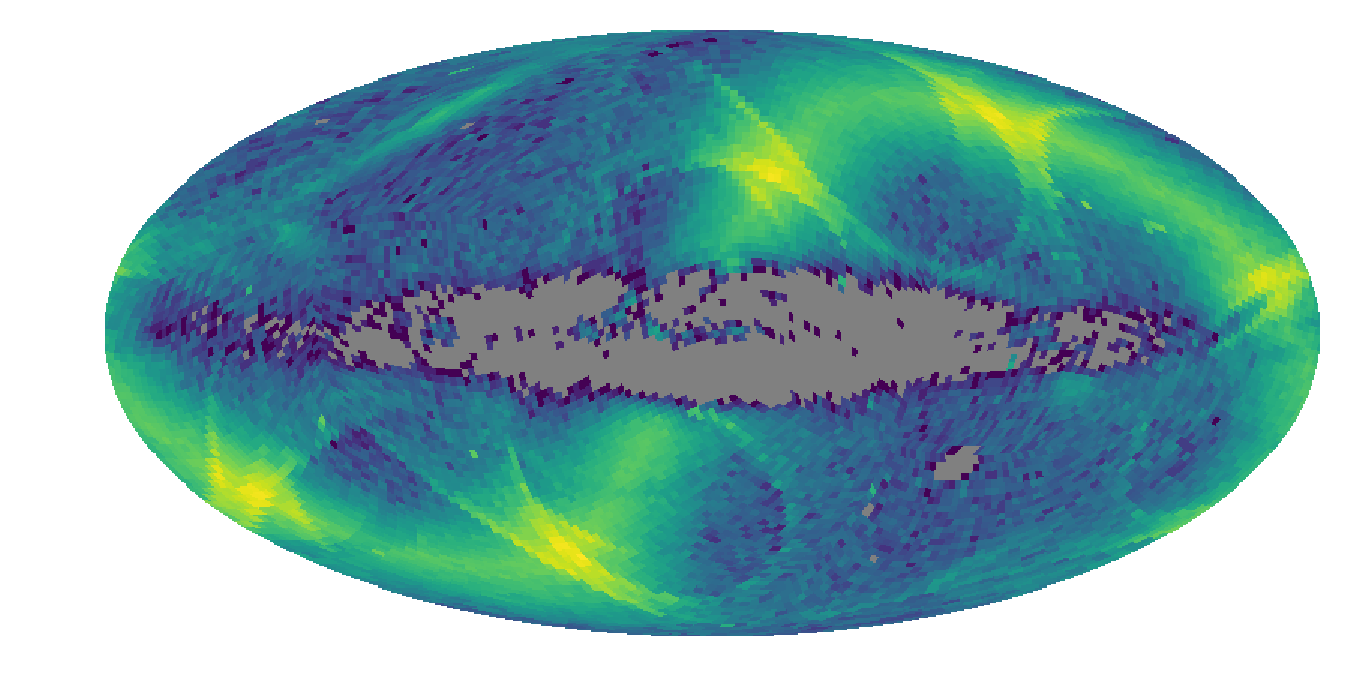}} 

\subfloat[]{\includegraphics[width=0.9\columnwidth]{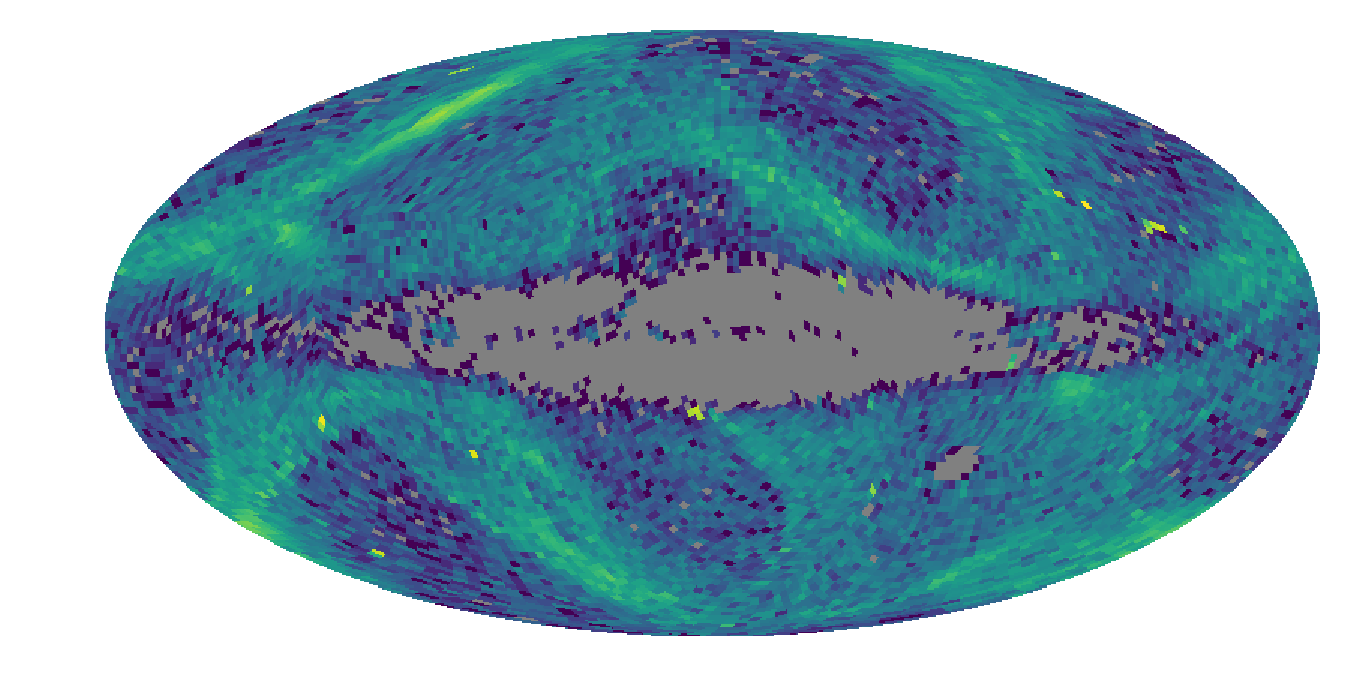}} 

\caption{The all-sky distribution of candidate transients. The map resolution is HEALPix of nside 32. The plots are in Galactic coordinates. (a) Almost 38 million new sources with at least 8 of the 9(8) AF CCD measurements were flagged by IDT during the year 2018. The artefacts from the scanning law and astrometric excursions are clearly visible. (b) The most dense regions and observations during astrometric excursions are removed (steps (ii) and (iii)). (c) Steps (iv-viii) are applied. The artefacts from the scanning law remain, however the excess in the number of sources along the ecliptic is visible. (d) The final map.}
\label{fig:mapbysteps}
\end{figure}

The impact of each selection criterion applied during filtering on the sample size is summarised in Tab. \ref{tab:cri}. The criteria (ii)-(v) and (ix) have the largest impact on the number of detected sources. In Fig. \ref{fig:mapbysteps} we also present the change in the sky distribution of ''new sources'' candidate transients when filters are applied.

\begin{table}
\centering
\caption{A summary of the impact of each selection criterion applied during filtering on the sample size.}
\begin{tabular}{l l l}
\hline
Criterion & \# of remaining candidates & Rejection ratio \\
\hline
(i) & $38\times10^6$ & 0.74\\ %37947318 \\
(ii) & $8.3\times10^6$ & 0.78 \\ %8270424 \\
(iii) & $7.0\times10^6$ & 0.16 \\ %6981427  \\
(iv) & $3.6\times10^6$ & 0.49 \\ %3567092  \\
(v) & $1.4\times10^6$ & 0.61 \\ %1446551  \\
(vi) & $1.2\times10^6$ & 0.14 \\ %1158468  \\
(vii) & $1.2\times10^6$ & $<$0.01 \\ %1150392  \\
(viii) & $9.3\times10^5$ & 0.19 \\ %  \\
(ix) & $1.3\times10^5$ & 0.86 \\ %  \\
\hline
\end{tabular}
\label{tab:cri}
\end{table}

\subsection{Analysis of detected candidates}

After applying all filters described above we obtained $0.13$M candidate new sources from one year of {\it Gaia} observations. However, some of the candidate transients could still be false positives as not all spurious detections will be filtered by our criteria above despite our best efforts.  E.g.~ local dense regions like nearby galaxies or star clusters will not be filtered out and might cause spurious detections when {\it Gaia} scans over these regions from different angles. Moreover, in regions of high source density new detections of real existing sources still happen due to resource limitations (priority on-board reading sources and limitation on data transfer to Earth, see \citealt{2016A&A...595A...1G}).

% individual transits 252882 v
% unique sourceids 244146 v
% good times 349.64d v
% good HP12 159,649,824 from 201,326,592 v
% rate: 244146./((159649824./201326592.)*4.*(180.*180./pi)*349.64) = 0.021 v
From the one year all-sky test we obtained about $\sim 0.13$M new candidate transients that gives us the transient rate about $\lesssim0.010$ per sq deg per day. As {\it Gaia} observes about 1000 sq deg per day we should detect about 10 new candidate transients every day.
Several studies tried to address the question how many transients (of the Galactic origin - novae, M-flares and extragalactic - mostly SNe, but also QSO-flares) should be detectable in an optical magnitude limited search. For example,  \cite{2019MNRAS.484.4507V} obtained rates for several types of transients. They focused on discovery of extragalactic optical fast transients and provided rates for transients faster than 1 day ($\lesssim 37\cdot10^{-4}$ per sq deg per d) and faster than 4 h ($\lesssim 9.3\cdot10^{-4}$ per sq deg per d with a limiting magnitude of $R\approx19.7$). In principle, {\it Gaia} should be able to detect many of these fast transients. \cite{2013ApJ...779...18B} identified at least two sources of potential false positives (M-star flares and asteroids) also relevant to searches for EM counterparts to GW events. The rates for other types of transients provided in \cite{2019MNRAS.484.4507V} ($\sim12\cdot10^{-4}$ SNe per sq deg per d, $\lesssim 20\cdot10^{-4}$ novae per sq deg per d and $\lesssim 120\cdot10^{-4}$ M-star flares per sq deg per d for a survey limited to $R<20$ mag) might give an estimate of how many false positive candidates are still included in the sample coming from our algorithm. 

To study candidate transients from our one year search we extended the period by one month before and after assuming that {\it Gaia} scans each part of the sky on average every 30 days, hence transients detected earlier (or later) by other surveys might be also detected by {\it Gaia} in 2018. According to the Transient Name Server (TNS; \url{https://wis-tns.weizmann.ac.il/}) during the period from 2017 December 01 to 2019 January 31, 10459 transients were recorded. We rediscovered 2948 transients in our one year long sky survey reported to TNS by various surveys (including 1289 alerted by GSA). In 2018 GSA discovered 2743 candidate transients where 1885 of them were found by the {\it NewSource} detector (although 485 of them are located in the dense regions excluded from our search and a few were initially too faint for our detector, hence the remaining 1400 transients should be potentially rediscovered here). In total 111 transients (8 per cent from the GSA sample) were not rediscovered by our detection algorithm even though they were discovered by GSA (these candidates were filtered out because of several different reasons, mostly due to their proximity to bright sources in GDR2 and inaccurate database entries regarding the IDT classification for new sources). 
All cross matches were performed using a search radius of 1 arcsec.

\section{Events from runs O1 and O2}
\label{sec:O1O2}

Each GW detection comes with a sky localisation map obtained from LIGO or LIGO-Virgo observations. Here, we made use of the final maps from the Gravitational-Wave Transient Catalog (\url{https://www.gw-openscience.org/GWTC-1-confident/}, \citealt{2019PhRvX...9c1040A}) for the events from the O1 and O2 runs. In total 11 events were found after reanalysis of the GW observations during these runs.

We used the {\it Gaia} Observation Forecast Tool (GOST)\footnote{\url{https://gaia.esac.esa.int/gost/} The GOST only provides a forecast of the time when targets cross the {\it Gaia} Focal Plane based on the scanning law of {\it Gaia}. However, it does not take into account operational activities preventing nominal observations nor the gaps between CCDs on the Focal Plane. Hence, the real number of scans may differ from the predictions.} to obtain a forecast for the visibility of each sky localisation map with 90 per cent credible regions. Assuming that the {\it Gaia} detection window of any putative EM counterpart to a GW event is short (time scale of days) we checked the visibility within 7 days from the GW events (see Table \ref{tab:gaia4o1o2}). For the three events detected during O1 (the BHBH mergers GW150914, GW151012, GW151226) the 90 per cent probability regions were partially scanned by {\it Gaia} within 7 days after the event. The fraction of the sky localisation regions scanned varies from 7 to 55 percent. For the five events detected during O2 (the BHBH mergers: GW170608, GW170809, GW170814, GW170818, and the binary NS merger GW170817) the 90 per cent probability regions were not scanned by {\it Gaia} within 7 days after the event time. The regions for the remaining three BHBH events (GW170104, GW170809, GW170823) were partially scanned (from 3 to 24 per cent). As one can expect the probability of {\it Gaia} observations increases with the size of the sky localisation map.
   
In Table \ref{tab:gaia4o1o2} we also included the predictions of the median time span between {\it Gaia} scanning over part of the sky localisation regions and the occurrence of the GW events, and the median time span between previous {\it Gaia} scans and the time of GW events. These time spans are strongly related to the localisation and size of the GW sky maps, and the uneven {\it Gaia} scanning law. Moreover, the minimal time delay between the GW event and {\it Gaia} observations in the sky localisation region might be lower than 0.01 d and as large as dozens of days.

\begin{table*}
\centering
\caption{The {\it Gaia} scanning predictions for the GW events from the LIGO (and Virgo) O1 and O2 runs.}
\begin{tabular}{l c c c c c c}
\hline
GW ID & $\Delta T$ [d] & Min $\Delta T$ [d] & Max $\Delta T$ [d] & $\Delta T_0$ [d] & $p$ [\%] & $\Delta \Omega~\mathrm{[deg^2]}$\\
\hline
GW150914
& 6.44 & 0.01 & 68.14 & -30.22 & 55 & 99 \\
GW151012
& 12.45 & 0.05 & 208.52 & -20.59 & 16 & 249 \\
GW151226
& 20.40 & 1.89 & 144.37 & -72.82 & 7 & 72 \\
GW170104
& 72.72 & $<$0.01 & 174.81 & -11.19 & 3 & 28 \\
GW170608
& 110.65 & 38.09 & 164.65 & -21.11 & 0 & - \\
GW170729
& 55.75 & $<$0.01 & 108.75 & -71.82 & 18 & 186 \\
GW170809
& 109.14 & 33.14 & 252.82 & -15.12 & 0 & - \\
GW170814
& 26.56 & 23.80 & 29.38 & -11.86 & 0 & - \\
GW170817
& 112.12 & 11.28 & 162.60 & -28.69 & 0 & - \\
GW170818
& 107.44 & 62.47 & 159.17 & -37.38 & 0 & - \\
GW170823
& 25.13 & 0.03 & 191.94 & -24.64 & 24 & 396 \\
\hline
\end{tabular}
\label{tab:gaia4o1o2}

$\Delta T$ - median wait time for {\it Gaia} to scan within the 90 per cent confidence sky localisation region of the GW event after the events, Min $\Delta T$, Max $\Delta T$ - minimum and maximum time delay between a GW event and {\it Gaia} to scan within the GW sky localisation, $\Delta T_0$ - median time between {\it Gaia} scanning in the sky localisation before the events and an occurring of the events, $p$ - percentage of a 90 per cent probability area scanned within 7 days from the event, $\Delta \Omega$ - size of scanned region within 7 days from the event
\end{table*}

For events covered by {\it Gaia} observations we ran a search for all candidate transients detected within 1 week after the GW events using the procedure described in Section \ref{sec:detector}. However, for events from 2015, due to the lack of internal information for SSO positions for {\it Gaia}, we needed to remove the candidates likely coinciding with SSO observations using their position in the sky with respect to the ecliptic. We also noticed that the sample is still affected by artefacts caused by bright stars (located further than 30 arcsec from the candidates - this is a tail of the neighbour distribution that was not taken into account in the criterion (v) due to a limited search radius). We obtain a sample of 250 candidates which were then visually inspected. 
The table \ref{tab:candO1O2} in the Appendix presents the candidate transients that pass our final eyeballing vetting (we provide coordinates, the discovery date, and the discovery magnitude). In addition, we identified candidates located on top of galaxies or in the vicinity of extended objects (about 40 per cent of candidates).

We made an attempt to study the completeness of our search by comparing the final sample of candidates with samples published on the TNS and the Gamma-ray Coordinates Network (GCN) circulars (\url{https://gcn.gsfc.nasa.gov/}). 
There are no transients alerted over the period covering the second part of 2015 and early 2016 as GSA was switched off over this period due to upgrades and testing of the current GSA detection algorithm. 
From all transients alerted by GSA after the GW events from the O2 run a single candidate was inside the 90 per cent probability contour of GW170823 (Gaia17cdt) and this transient was rediscovered using the method outlined above. Moreover, two other sources reported to TNS by the PS1 survey were rediscovered in our work (AT2017jxq, AT2017gpc - both sources were too faint to be discovered by the current GSA {\it NewSource} detector). A few other published transients were detected just before the events (i.e.~Gaia17aba, Gaia17aaw were discovered within 2 days before GW170104, Gaia17bxj was found within 0.15 d before GW170729, and Gaia17cct within 3 days before GW170823). These sources are useful to assess results from other surveys where a search for EM counterparts usually starts after the detection of a GW event.

Unfortunately, {\it Gaia} was not scanning the region of the sky localisation for the GW170817 event during and shortly after the event. From the observations about 2 weeks after the GW signal we only know that no source was detected and that is consistent with the detection limits from other survey (e.g.~PS1 upper limits 12 days after the kilonova peak are $g>22.5$ mag and
 $r>21.7$ mag, \citealt{2017Natur.551...75S}).
 
\subsection{Candidates from existing sources}

Candidate transients can also be detected as a change in brightness of a source detected by {\it Gaia} before the GW event. This can for instance happen if the location of a transient cannot be spatially separated from its host galaxy. For GW event counterparts this can only happen if the event is typically closer to the centre of host galaxy than 1-2 arcsec or if the host has a small angular size (i.e. it is a dwarf galaxy or a galaxy located at higher redshift). 
For all events from the O1 and O2 run we searched for {\it Gaia} transits occurring within 7 days from GW detection. We required that the transits are not flagged as a new source by IDT. This criterion implies that the transit is associated with a source existing in the database of sources detected by {\it  Gaia} before the GW event. We required 8-9 valid AF measurements, a median flux above $101.5~\mathrm{e^{-}/s}$, and restricted scatter between AF measurements during the candidate transit. These criteria still leave about 15M transits for which we tried to build light curves using data from previous {\it Gaia} scans collected within a search radius of 0.5 arcsec from the candidate transits. Furthermore, all sources were cross matched with the GLADE catalogue (Galaxy List for the Advanced Detector Era, a full sky galaxy catalogue, \citealt{2018MNRAS.479.2374D}) with a search radius of 1 arcsec to obtain candidates with a location close to or consistent with the centres of (known) galaxies. We limited the sample to sources with a flux increase after the GW event of more than $5\sigma$ when compared to the median flux detected by {\it Gaia} before the GW event. Forty eight candidates were found. Forty one candidates were classified as false positives after manual vetting. These false positives are mostly related to candidates just above the detection threshold, these could be caused by flux variations due to changes in {\it Gaia}'s scan angle over the source. There are a few exceptions which we classify as bright stars that are mislabelled as galaxy in the GLADE catalogue. We are left with seven transient candidate (including Gaia18cqe - a transient that was detected and alerted through the existing GSA system and for which it is known that it is related to known blazar activity, see \citealt{2017ATel10482....1C}). Our remaining six candidates also show evidence of being caused by quasar activity. Interestingly, some of these have a redshift that puts them at a distance within $<1-3\sigma$ of the distance of the observed GW events. 

\subsection{Future prospects}
\label{sec:dis}

\begin{figure}
\includegraphics[width=\columnwidth]{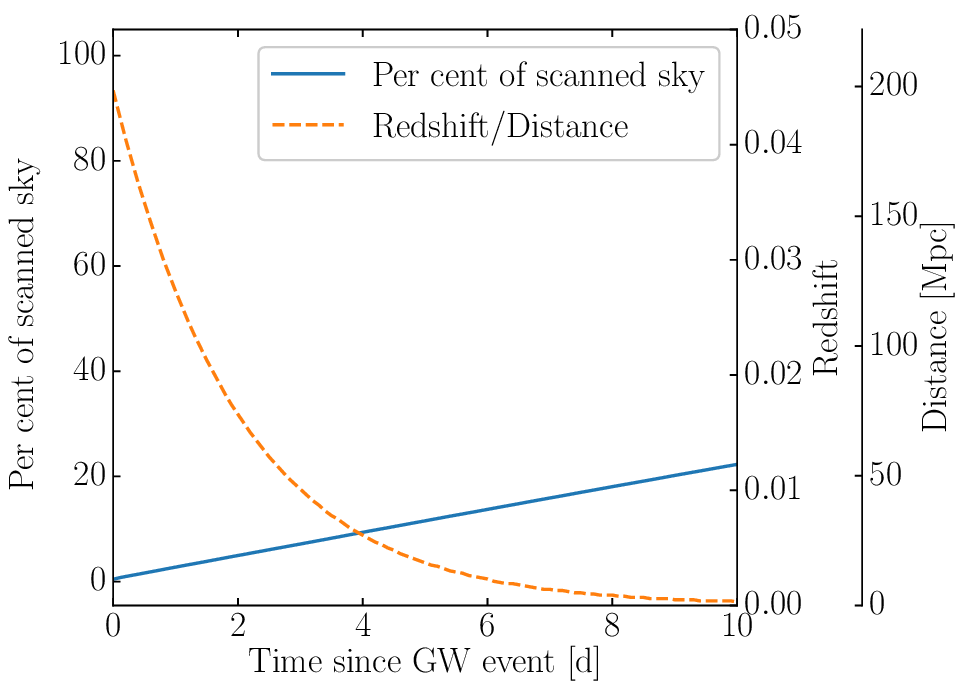}
\caption{Redshift (distance) limits for any {\it Gaia} detection of kilonovae (orange dashed line), and per cent of scanned sky by {\it Gaia} (blue line) vs.~time since a GW event. Assuming a model of kilonovae similar to the transient detected as GW170817/AT2017gfo (e.g.~\citealt{2017Natur.551...75S}) we predict that {\it Gaia} will be able to detect such transients from a redshift up to $\sim 0.045$. However, the sample of detected events might be limited by {\it Gaia}'s scanning law as only about 25 per cent of the sky will be scanned within 10 days after a GW event.}
\label{fig:knLimits}
\end{figure}

We studied chances of future EM counterpart detections by {\it Gaia}. Assuming that the kilonova events have an absolute magnitude about -15.8 mag in $r$-band at peak (i.e.~similar to the first GW event--kilonova source GW170817/AT2017gfo (e.g. \citealt{2017Natur.551...75S}) and our {\it Gaia} detection threshold of $\sim 20.68$ mag in $G$-band, transients can be detected up to redshift $\sim 0.045$ (see Fig.~\ref{fig:knLimits}). Within 10 days from a GW event about 25 per cent of sources randomly located in the sky will be in the regions scanned by {\it Gaia}. Although, detections of individual events are strictly related to localisation as {\it Gaia} scanning law is uneven over the sky.

An EM counterpart to the GW event might be also discovered by the {\it OldSource} detector where the host galaxy was previously detected by {\it Gaia} and a candidate transient is not resolved from its host. Over the year 2018 from all 2743 published {\it Gaia} transients more than 30 per cent was found in this way. However, in this sample there are also included candidates for microlensing events, cataclysmic variables, AGN flares, star flares, young stellar objects.

\section{Conclusions}
\label{sec:con}

We studied the possibility of detecting EM counterparts to GW events by {\it Gaia} using a dedicated transient detection algorithm. We propose an extension to current algorithms used by GSA to find transients and tested its capabilities in discovering candidate transients. The main concern is the level of false positives which has to be limited through various filters (based on event time, sky localisation map, but also candidate transient neighbourhood). 
The candidate transients for EM counterparts to the previous GW detections are also reported. The search using the bespoke detector yielded 250 candidate transients observed by {\it Gaia} within 7 days from the GW events detected in the O1 and O2 runs.
As GSA were not publishing new transients between July 2015 and January 2016 for GW events from the O1 run the sample of candidates was only compared to results from other surveys. For candidates from the year 2017 (run O2) one candidate transient was alerted by GSA and rediscovered by this search. Moreover, we rediscovered two transients from other surveys reported to TNS.
For the GW events from the current O3 run we expect that about 16 (25) per cent of them might be in the sky regions observed by {\it Gaia} within 7 (10) days from the events. The new detector will provide $\lesssim 10$ candidates per day from the whole sky (the final number of candidates will be lower as the GW localisation skymap usually are the size of a few thousand square degrees). One of the main advantages of using {\it Gaia} in study transients associated with GW signal is the accurate position (up to mas). Moreover, thanks to {\it Gaia} position on the orbit we are able to observe targets relatively close to the Sun, areas that are not easily reachable from ground based observatories.

\section*{Acknowledgements}
ZKR acknowledges funding from the Netherlands Research School for Astronomy (NOVA).
ZKR, PGJ, and DE acknowledge support from European Research Council Consolidator Grant 647208.
{\L}W acknowledges Polish NCN HARMONIA grant No. 2018/30/M/ST9/00311.
TW is funded in part by European Research Council grant 320360 and by European Commission grant 730980.
This publication is based upon work from COST Action MW-Gaia CA18104 supported by COST (European Cooperation in Science and Technology).

This work has made use of data from the European Space Agency (ESA)
mission {\it Gaia} (\url{https://www.cosmos.esa.int/gaia}), processed by
the {\it Gaia} Data Processing and Analysis Consortium (DPAC,
\url{https://www.cosmos.esa.int/web/gaia/dpac/consortium}). Funding
for the DPAC has been provided by national institutions, in particular
the institutions participating in the {\it Gaia} Multilateral Agreement.

Funding for the Sloan Digital Sky Survey IV has been provided by the Alfred P. Sloan Foundation, the U.S. Department of Energy Office of Science, and the Participating Institutions. SDSS-IV acknowledges
support and resources from the Center for High-Performance Computing at
the University of Utah. The SDSS web site is www.sdss.org. SDSS-IV is managed by the Astrophysical Research Consortium for the 
Participating Institutions of the SDSS Collaboration including the 
Brazilian Participation Group, the Carnegie Institution for Science, 
Carnegie Mellon University, the Chilean Participation Group, the French Participation Group, Harvard-Smithsonian Center for Astrophysics, 
Instituto de Astrof\'isica de Canarias, The Johns Hopkins University, Kavli Institute for the Physics and Mathematics of the Universe (IPMU) / 
University of Tokyo, the Korean Participation Group, Lawrence Berkeley National Laboratory, 
Leibniz Institut f\"ur Astrophysik Potsdam (AIP),  
Max-Planck-Institut f\"ur Astronomie (MPIA Heidelberg), 
Max-Planck-Institut f\"ur Astrophysik (MPA Garching), 
Max-Planck-Institut f\"ur Extraterrestrische Physik (MPE), 
National Astronomical Observatories of China, New Mexico State University, 
New York University, University of Notre Dame, 
Observat\'ario Nacional / MCTI, The Ohio State University, 
Pennsylvania State University, Shanghai Astronomical Observatory, 
United Kingdom Participation Group,
Universidad Nacional Aut\'onoma de M\'exico, University of Arizona, 
University of Colorado Boulder, University of Oxford, University of Portsmouth, 
University of Utah, University of Virginia, University of Washington, University of Wisconsin, 
Vanderbilt University, and Yale University.

The Pan-STARRS1 Surveys (PS1) and the PS1 public science archive have been made possible through contributions by the Institute for Astronomy, the University of Hawaii, the Pan-STARRS Project Office, the Max-Planck Society and its participating institutes, the Max Planck Institute for Astronomy, Heidelberg and the Max Planck Institute for Extraterrestrial Physics, Garching, The Johns Hopkins University, Durham University, the University of Edinburgh, the Queen's University Belfast, the Harvard-Smithsonian Center for Astrophysics, the Las Cumbres Observatory Global Telescope Network Incorporated, the National Central University of Taiwan, the Space Telescope Science Institute, the National Aeronautics and Space Administration under Grant No. NNX08AR22G issued through the Planetary Science Division of the NASA Science Mission Directorate, the National Science Foundation Grant No. AST-1238877, the University of Maryland, Eotvos Lorand University (ELTE), the Los Alamos National Laboratory, and the Gordon and Betty Moore Foundation.

This research has made use of the SIMBAD database,
operated at CDS, Strasbourg, France.

This research has made use of Astropy, a community-developed core Python package for Astronomy \citep{2013A&A...558A..33A}, healpy, a Python package to manipulate HEALPix maps (\url{http://healpix.sf.net}, \citealt{2005ApJ...622..759G,Zonca2019}), Q3C extension for PostgreSQL \citep{2006ASPC..351..735K}, \textsc{TOPCAT} \citep{2005ASPC..347...29T}.

%%%%%%%%%%%%%%%%%%%%%%%%%%%%%%%%%%%%%%%%%%%%%%%%%%



\bibliographystyle{mnras}
\bibliography{emgw}

%%%%%%%%%%%%%%%%%%%%%%%%%%%%%%%%%%%%%%%%%%%%%%%%%%

%%%%%%%%%%%%%%%%% APPENDICES %%%%%%%%%%%%%%%%%%%%%

\appendix

\section{Candidate transients for O1 and O2 events}

Here we present 250 candidate transients associated with GW events from the O1 and O2 runs detected using {\it Gaia}. The table contains an overview of all transients sorted by GW events and discovery date (coordinates, discovery date, discovery magnitude, nearby potential hosts if known). The discoveries by other surveys are also mentioned (where applicable). The potential hosts were identified by a cross-match using the GLADE catalogue (\citealt{2018MNRAS.479.2374D}), the SDSS catalogue  (\citealt{2017AJ....154...28B}), and the SIMBAD database (\citealt{2000A&AS..143....9W}) within a search radius of 15 arcsec.

\begin{table*}
\centering
\caption{Candidate transients from {\it Gaia}. Note: This table is available in its entirety in a machine-readable form from the online journal. A portion is shown here for guidance regarding its form and content.}
\begin{tabular}{c c c c c}
\hline
RA & Dec & JD & Mag & Comments \\
Deg & Deg & TCB & {\it Gaia}-$G$ & host + other surveys\\
\hline
\multicolumn{5}{c}{GW150914} \\
\hline
57.775673 & -59.139924 & 2457279.920880 & 19.172 $\pm$ 0.016 & 03510664-5908196 \\
53.595064 & -61.599607 & 2457280.349236 & 20.401 $\pm$ 0.171 &  \\
62.315471 & -62.761413 & 2457281.171551 & 20.087 $\pm$ 0.032 & 2MASXJ04091581-6245364 \\
58.805149 & -64.886501 & 2457281.423345 & 19.529 $\pm$ 0.043 &  \\
58.011188 & -64.991036 & 2457281.423588 & 20.213 $\pm$ 0.041 & 03520183-6459230 \\
\hline
\multicolumn{5}{c}{GW151012} \\
\hline
67.105278 & 50.180461 & 2457313.753935 & 19.187 $\pm$ 0.048 &  \\
67.101160 & 50.185544 & 2457313.827928 & 19.234 $\pm$ 0.047 &  \\
\hline
\multicolumn{5}{c}{GW151226}  \\
\hline
192.155376 & 7.457991 & 2457388.042998 & 18.459 $\pm$ 0.013 & SDSSJ124837.59+072723.3 \\
192.033382 & 7.271310 & 2457388.043125 & 20.195 $\pm$ 0.097 &  \\
191.528950 & 7.530792 & 2457388.043380 & 19.783 $\pm$ 0.028 &  \\
\hline

\end{tabular}
\label{tab:candO1O2}
\end{table*}

%%%%%%%%%%%%%%%%%%%%%%%%%%%%%%%%%%%%%%%%%%%%%%%%%%

% Don't change these lines
\bsp	% typesetting comment
\label{lastpage}
\end{document}